\documentclass[onecolumn]{aastex631}
\usepackage[utf8]{inputenc}
\usepackage{multirow}
\usepackage[numbib]{tocbibind}
\usepackage{amsmath,amsthm,amssymb}
\usepackage{longtable}
\usepackage[T1]{fontenc}
\usepackage{threeparttable}

\begin{document}

\title{Functionality of Ice Line Latitudinal EBM Tenacity (FILLET). Protocol Version 1.0.\\
A CUISINES intercomparison project.}
\author[0000-0001-9423-8121]{Russell Deitrick}
%\affil{Center for Space and Habitability, University of Bern, Gesellschaftsstrasse 6, CH-3012, Bern, Switzerland}
\affil{School of Earth and Ocean Sciences, University of Victoria, Victoria, British Columbia, Canada}
\email{rdeitrick@uvic.ca}
\author[0000-0003-4346-2611]{Jacob Haqq-Misra}
\affil{Blue Marble Space Institute of Science, Seattle, WA, USA}
\author[0000-0002-5826-1540]{Shintaro Kadoya}
\affil{Japan Agency for Marine-Earth Science and Technology, X-star, Kanagawa, Japan}
\author[0000-0001-7553-8444]{Ramses Ramirez}
\affil{University of Central Florida, Department of Physics, Planetary Sciences Group, Orlando, Fl. 32816}
\author[0000-0002-7744-5804]{Paolo Simonetti}
\affiliation{University of Trieste, Dep. of Physics, Via G.~B. Tiepolo 11, I-34143 Trieste, Italy}
\affiliation{INAF Trieste Astronomical Observatory, Via G.~B. Tiepolo 11, I-34143 Trieste, Italy}
\author[0000-0001-6487-5445]{Rory Barnes}
\affil{Astronomy Department, University of Washington, Seattle, WA, USA 98105-1580}
\author[0000-0002-5967-9631]{Thomas J. Fauchez}
\affiliation{NASA Goddard Space Flight Center
8800 Greenbelt Road
Greenbelt, MD 20771, USA}
\affiliation{Integrated Space Science and Technology Institute, Department of Physics, American University, Washington DC}
\affiliation{NASA GSFC Sellers Exoplanet Environments Collaboration}
% \author[0000-0002-6673-2007]{Linda Sohl}
% \affiliation{Center for Climate Systems Research, Columbia University, New York, NY, USA}
% \affiliation{NASA Goddard Institute for Space Studies, 2880 Broadway, New York, NY 10025, USA}
\date{}

% \linenumbers
\begin{abstract}
Energy balance models (EBMs) are one- or two-dimensional climate models that can provide insight into planetary atmospheres, particularly with regard to habitability. Because EBMs are far less computationally intensive than three-dimensional general circulation models (GCMs), they can be run over large, uncertain parameter spaces and can be used to explore long-period phenomena like carbon and Milankovitch cycles. Because horizontal dimensions are incorporated in EBMs, they can explore processes that are beyond the reach of one-dimensional radiative-convective models (RCMs). EBMs are, however, dependent on parameterizations and tunings to account for physical processes that are neglected. Thus, EBMs rely on observations and results from GCMs and RCMs. Different EBMs have included a wide range of parameterizations (for albedo, radiation, and heat diffusion) and additional physics, such as carbon cycling and ice sheets. This CUISINES exoplanet model intercomparison project (exoMIP) will compare various EBMs across a set of numerical experiments.  The set of experiments will include Earth-like planets at different obliquities, parameter sweeps across obliquity, and variations in instellation and CO$_2$ abundance to produce hysteresis diagrams.  We expect a range of different results due to the choices made in the various codes, highlighting which results are robust across models and which are dependent on parameterizations or other modeling choices. Additionally, it will allow developers to identify model defects and determine which parameterizations are most useful or relevant to the problem of interest. Ultimately, this exoMIP will allow us to improve the consistency between EBMs and accelerate the process of discovering habitable exoplanets.
\end{abstract}

\section{Introduction}
Computational climate models enable the exploration of planetary climate over broad parameter spaces, which can be useful for understanding the evolution of Earth's climate and the diversity of exoplanet atmospheres. Energy balance models (EBMs) consider the radiative balance between incoming stellar radiation and outgoing infrared radiation as the primary driver of climate. Like most climate models, EBMs were first developed for understanding climate on Earth. The first EBMs were developed to examine the glacial cycles of the last $\sim$5 Myr \citep{budyko1969} and the potential response of ice cap loss due to global warming \citep{sellers1969}. These pioneering studies predicted that Earth's climate is bistable between glacial and warm states, and subsequent works expanded the capabilities of such models to elucidate the role of ice-albedo feedback on climate stability \citep[e.g.][]{northcoakley1979,north1981energy}. The use of EBMs facilitated the discovery of the small ice cap instability (SICI) and large ice cap instability (LICI) \citep{cahalan1979}. EBMs can include numerous parameterizations to account for seasonal cycles, meridional energy transport, longitudinal variation, greenhouse gas forcing, ocean feedback, and other processes; see \citet{north2017} for an extensive discussion on using EBMs to study present-day and ancient climates. EBMs have also been applied to study the climate history of Mars \citep[e.g.][]{lorenz2001, armstrong2004, fairen2012, batalha2016, hayworth2020}. 

Recently, EBMs have had a revival in the domain of exoplanets, owing to the large uncertainties in planet properties and the computational load of other climate models. The first application to hypothetical exoplanets was done in \cite{williamskasting97}. A number of subsequent studies have been used to study potentially habitable exoplanets with diverse orbits and rotational properties \citep{spiegel2008,spiegel2009,spiegel2010,dressing2010,armstrong2014,forgan2014,forgan2016,may2016,checlair2017,rose2017,silva2017,deitrick2018,okuya2019,haqqmisra2019,yadavalli2020,palubski2020,wilhelm2022}. Others have focused on atmospheric or surface properties  \citep{shields2013,vladilo2013,haqqmisra2014,kadoya2014,menou2015,haqqmisra2016,kadoya2016,kadoya2019,rushby2019,ramirez2018ice,ramirez2020,bonati2021,haqqmisra2022}. Thus a rich body of literature applying EBMs to exoplanets exists. 

Only a few climate modeling intercomparisons of exoplanets have been performed to date, and only for GCMs \citep{yang2019,fauchez2020}. These studies drew inspiration from the intercomparison projects of the Earth sciences community to better quantify the effects of global warming \citep{eyring2016}. To our knowledge, no such comparison has been conducted for EBMs, either for Earth or any other planet. This is perhaps because of the relative simplicity of the EBM compared to GCMs. However, given the absence of exoplanetary data, and the increasing dependence of exoplanet EBMs on various parameterizations, an intercomparison of frequently used models is warranted and timely. Here, we propose the first intercomparison of exoplanet EBMs, FILLET (Functionality of Ice Line Latitudinal EBM Tenacity). This is one facet of the international CUISINES (Climates Using Interactive Suites of Intercomparisons Nested for Exoplanet Studies) model intercomparison framework for exoplanets, which began with THAI (TRAPPIST Habitable Atmosphere Intercomparison) \citep{fauchez2020}. Participation in FILLET, like the other CUISINES projects, is open to any researcher with access to  a suitable model. 

This article is organized as follows. Section 2 presents basic information on the EBM and our motivation for this intercomparison project. Section 3 describes the numerical experiments to be performed by each participating model. Section 4 describes the outputs and data formats that each model will produce. Finally, Section 5 presents a brief summary. 

\section{Motivation/goals}

The back-bone of all 1-D or 2-D horizontal\footnote{That is, the one dimension is either latitude or longitude, while 2-D models have both latitude and longitude.} energy balance models is a single partial differential equation describing the temperature evolution of different locations on the planet,
\begin{equation}
    C \frac{\partial T}{\partial t} - \nabla \cdot \kappa \nabla T + I = S (1-\alpha), \label{eqn:ebm}
\end{equation}
where $T$ is the surface temperature, $C$ is the surface heat capacity by area, $\kappa$ is thermal conductivity, $I$ is the outgoing long-wave radiation (OLR), $S$ is the incident stellar flux (instellation), and $\alpha$ is the albedo (alternatively defined at the surface or the top of the atmosphere). Many one-dimensional EBMs use a constant diffusive parameter $D$, defined as $D = \kappa/R^2$, which then appears outside the partial derivative ($R$ is the planet radius). The instellation $S$ is determined purely by the orbit, obliquity, and location on the planet, and so can be computed exactly. This leaves four quantities, $C$, $D$, $I$, and $\alpha$, which must then be parameterized. The majority of the uncertainty in energy balance modeling lies in this parameterization process. Different EBMs use a diverse set of parameterizations for each property. 

Additionally, there are a number of studies that have coupled EBMs with additional physics models, such as sea ice \citep{shields2013}, weathering \citep{williamskasting97,kadoya2014,haqqmisra2014}, ice sheets \citep{huybers2008, deitrick2018}, and proposed carbon dioxide cycling mechanisms on ocean worlds \citep{ramirez2018ice}. The behavior of EBMs with these additional models has not been as well characterized as the uncoupled models. 

Results from different EBMs can vary widely (compare, e.g., \cite{armstrong2014} vs. \cite{deitrick2018}). Sometimes the reasons are clear, as in the aforementioned example---as explained in \citet{deitrick2018}, the results differ due to the different OLR prescriptions and the diffusion term, which was missing in \citet{armstrong2014}. The net effect is that the EBM used in \citet{armstrong2014} is much more stable against ice-albedo instabilities, perhaps unrealistically so. Often the sources of discrepancies are more difficult to pin-point. It would be compelling to identify features in EBM results that are more-or-less universal across models, such as equatorial superrotation is in GCM simulations of hot Jupiters \citep[see][]{heng2015}. It is also interesting to determine the distribution of results for certain properties across models, such as the spread of global mean surface temperatures in the THAI intercomparison project \citep{turbet2021,sergeev2021}, as this informs future modelers of the potential strengths or weaknesses of each model.

One of the central concepts of this intercomparison is that of an ``ice line'', or the average latitude to which ice caps or equatorial ice ``belts'' extend. In this exoMIP, we seek an understanding of the robustness or ``tenacity'' of the ice line across EBMs as well as an understanding of its functional dependencies. Of course, the definition of the ice line is model-dependent because some models incorporate fractional ice coverage at each latitude and each model has different ways of determining whether ice is present. Thus another objective of this exoMIP is to provide clear guidance on how to define the ice line. While most models since \citet{budyko1969} and \citet{sellers1969} have directly connected the ice line to temperature through a temperature dependent surface albedo, studies by \citet{walsh2016, walsh2020, nadeau2021} have shown the usefulness of representing ice sheets explicitly, thus creating an indirect temperature-albedo connection. One of the EBMs in FILLET (\texttt{VPLanet/POISE}) includes an explicit ice sheet model, thus will be capable of providing some additional insight on this issue in our experiments. However, as we describe in Section \ref{sec:protocol}, we are generally running models to steady-state, so the time-delay imposed by the ice sheet model may be largely irrelevant.

The goals of this intercomparison project are to: (1) quantify the effects of various parameterizations used in the participating EBMs; (2) identify results that are robust across all models; (3) set benchmarks for future EBM development; (4) quantify the effects of any additional physics incorporated in the EBMs; and (5) isolate any existing coding errors in the participating models. 

A list of EBMs active in exoplanet science is presented in Table \ref{tab:models}; most of these are already participating in FILLET. This list is not necessarily complete. Researchers who are interested in participating with their own EBM, even if it shares a coding lineage with a model listed here, are invited to contact the authors of this work to join FILLET. 

\begin{table*}[]
    \centering
      \caption{Currently active EBMs in exoplanet science}
    \begin{tabular}{lcc}
    \hline
    EBM & References & Notes\\
    \hline
       \texttt{HEXTOR}$^{\dagger}$ & \citet{haqqmisra2014,haqqmisra2022} & Successor to Darren Williams model \\
       \texttt{VPLanet/POISE}$^{\dagger}$   & \citet{deitrick2018,barnes2020}& Adapted from Cecilia Bitz model\\
       Kadoya-Tajika model$^\dagger$ & \citet{kadoya2014,kadoya2019} & Variation of Darren Williams model \\
       \texttt{OPS}$^{\dagger}$  & \citet{ramirez2020,ramirez2020complex} & Variation of Darren Williams model\\
       \texttt{PlaHab} & (Ramirez 2023, in prep.) & 2-D lat-lon EBM \\
       Shields-Bitz model & \citet{shields2013,rushby2019} & Variation of Cecilia Bitz model\\
       \texttt{ESTM}$^\dagger$ & \citet{vladilo2013,vladilo2015,biasiotti2022} & Meridional-vertical EBM\\
       \hline
    \end{tabular}\\
    $^{\dagger}$ Active participant in FILLET at time of submission
    \label{tab:models}
\end{table*}

\section{FILLET protocol}
\label{sec:protocol}
% Participants should run their models using nominal/published configurations where possible, unless alternate settings are needed for a particular comparison. The concept here is to perform ``unbiased'' tests of each code. After this, we can sort through differences and determine which are easily explained vs. which are potentially troubling. \rdcmt{Some models are able to perform each experiment with different settings---we should discuss whether these are worth doing.}

Participants should run their models using values for the spatial and time resolution that are suitable for the model and experiment. Because there are so many differences in how the models are discretized in time and space, these values may vary substantially. As an example, \texttt{VPLanet/POISE} can use any number of latitudes \citep[][used 150]{deitrick2018}, as can \texttt{OPS} and \texttt{ESTM}, while \texttt{HEXTOR} uses a fixed number of 18 latitudes. Some experimentation may be necessary to determine the required resolution of each experiment. Time steps in EBMs can vary from $\sim$10s of minutes to several days. It is sufficient for the  time steps to be small enough to achieve convergence (a steady mean state) and resolve seasonal effects---thus the time steps for high obliquity cases may be smaller than for low obliquity, for example. 

 Some models (e.g., \citet{williamskasting97} and successor models) have a built in annual mode, i.e., a numerical solution to the annual EBM equation wherein the time derivative in Equation \ref{eqn:ebm} is set to zero and the instellation is averaged over an orbit. The annual model has known analytical solutions \citep{cahalan1979,rose2017}, is readily benchmarked against those and other works, and is simple enough that intercomparison is not likely to yield much of interest. Furthermore, the annual and seasonal solution diverge substantially at moderate to high obliquity, indicating that the annual model is probably more limited in application \citep{rose2017}. Thus we will focus exclusively on the seasonal EBM (i.e., the solution to the full EBM equation) in this exoMIP.

As mentioned, each run will need to be evaluated for convergence, i.e., steady-state conditions. This can be done by examining multiple fields, such as global temperature, temperature at certain latitudes, etc., for steady state. While seasonal variations should persist, the annual averages should asymptotically approach a constant value. For a typical Earth-like case, this may be on the order of decades of integration time. There may, however, be additional physics that increase this time scale significantly. In \texttt{VPLanet/POISE}, for example, the inclusion of ice sheets can increase the convergence time to tens of thousands of years. EBMs incorporating carbon cycles may likewise require a longer convergence time, though for the first stage of the exoMIP we have no plans to model these cycles. 

The heat diffusion across the land-ocean boundary or across longitudes, if included as part of a two-dimensional EBM, is usually parameterized with a different value of the diffusion parameter, $D$, compared to the latitudinal component. This yields yet further tuning parameters. Models that have this additional parameter should begin with a standard (published) tuning. These tunings may require some adjustment, if results are significantly different.  

For this project, we will assume zero eccentricity to enforce hemispheric symmetry in the instellation. With zero eccentricity, the values of the precession angle (the azimuthal angle related to the spin axis) and longitude of pericenter (the orientation of the orbit's major axis) are irrelevant and thus do not need to be specified. 
%\textbf{Continuing phases of FILLET may explore a range of eccentricities, but these are not yet planned. The outcomes of the initial phase will inform follow-up phases (see Section \ref{sec:summary})}

Values for the instellation, CO$_2$ abundance, obliquity, and semi-major axis for the different cases are shown in Table \ref{tab:Params1}. The CO$_2$ abundance is given by the volume mixing ratio, which is related to the partial pressure via $X_{\text{CO}_2} = p_{\text{CO}_2}/p_{\rm total}$, where the total pressure is $p_{\rm total} = 1$ bar in all our cases. The semi-major axis is relevant here in that it determines the orbital period, and thus affects the length and strength of the seasons.

The primary effect of CO$_2$ abundance is to change the OLR. This is often accomplished via parameterizations or look-up tables derived from radiative-convective modeling outputs. An example of the latter is \texttt{ESTM}, which uses OLR and TOA albedo tables generated by \texttt{EOS} \citep{simonetti2022}. In some models (e.g. Williams model and successors), the CO$_2$ partial pressure is an input quantity. Some EBMs may use a linearized form of the OLR , $I = A + B T$ (e.g., \citet{caldeira1992}). The coefficients for a given value of $p_{\text{CO}_2}$ can be determined by using the polynomial fits from \cite{williamskasting97,haqqmisra2016,kadoya2019}, for example, and the following formulae:
\begin{align}
    B(p_{\text{CO}_2}) &= \frac{dI(p_{\text{CO}_2}, T)}{dT}\\
    A(p_{\text{CO}_2}) &= I(p_{\text{CO}_2}, T) - B(p_{\text{CO}_2})T,
\end{align}
where $T$ is a representative temperature (e.g., a rough guess of the expected mean temperature). Models participating in FILLET should use whichever model or parameterization for OLR and CO$_2$ is available in that model, with a preference for settings used in existing publications. Models that do not have a preexisting parameterization may use linear coefficients derived as described above or may request these coefficients from other FILLET participants (e.g., the authors of the present work).

Nominal values for surface albedo, heat diffusion, and heat capacity are identical across all cases, and are shown in Table \ref{tab:Params2}. Participants should use the constant values prescribed in this table, where possible. Other values or more complex parameterizations may be used \citep[e.g., the surface albedo in][]{williamskasting97} if using the prescribed values proves too challenging. For the top-of-atmosphere (TOA) albedo, participants should use the nominal configuration/parameterization for their model. 

The set of numerical experiments is described in detail below. Note that these may evolve as we begin comparisons, like some of the THAI simulations did. We propose three single simulation ``benchmarks'', which can be used to identify the most immediate issues, and four ``experiments'', which are more exploratory and will push the limits of parameterizations further. With each stage, in the event of sizable differences between models, we will collaborate to identify the source of each difference. 

With the exception of Benchmark 1, no tuning of the models should be done prior to the analysis and comparison stage. As a part of the analysis, we may choose to do tuning to bring the models into better agreement---this may be necessary to understand the source of model differences. To begin with, however, we will simply run the models using the prescribed settings and compare the resulting outputs. 

\subsection{Benchmark 1: Pre-industrial Earth}
This benchmark can be thought of as a ``sanity-check''---ensuring that each model can be configured to produce an Earth-like state. For this benchmark, we provide no prescriptions or guidance, other than that the CO$_2$ mixing ratio should be 280 ppm. Instead, modelers should perform their best attempt to produce a pre-industrial Earth, making whatever modeling choices or tunings they need. The aim is to achieve global mean surface temperature of 288 K. Outputs between models may still differ substantially in the details. Analysis of this case will be focused on the output variables and input parameters in equal measure. Example output for this benchmark from \texttt{VPLanet/POISE} is shown in Figure \ref{fig:earth} and output from \texttt{ESTM} is shown in Figure \ref{fig:earth_estm} and Table \ref{tab:estm_ben123}. 

\subsection{Benchmark 2: Un-tuned model, low obliquity}
\begin{figure*}
\centering
\includegraphics[width=\textwidth]{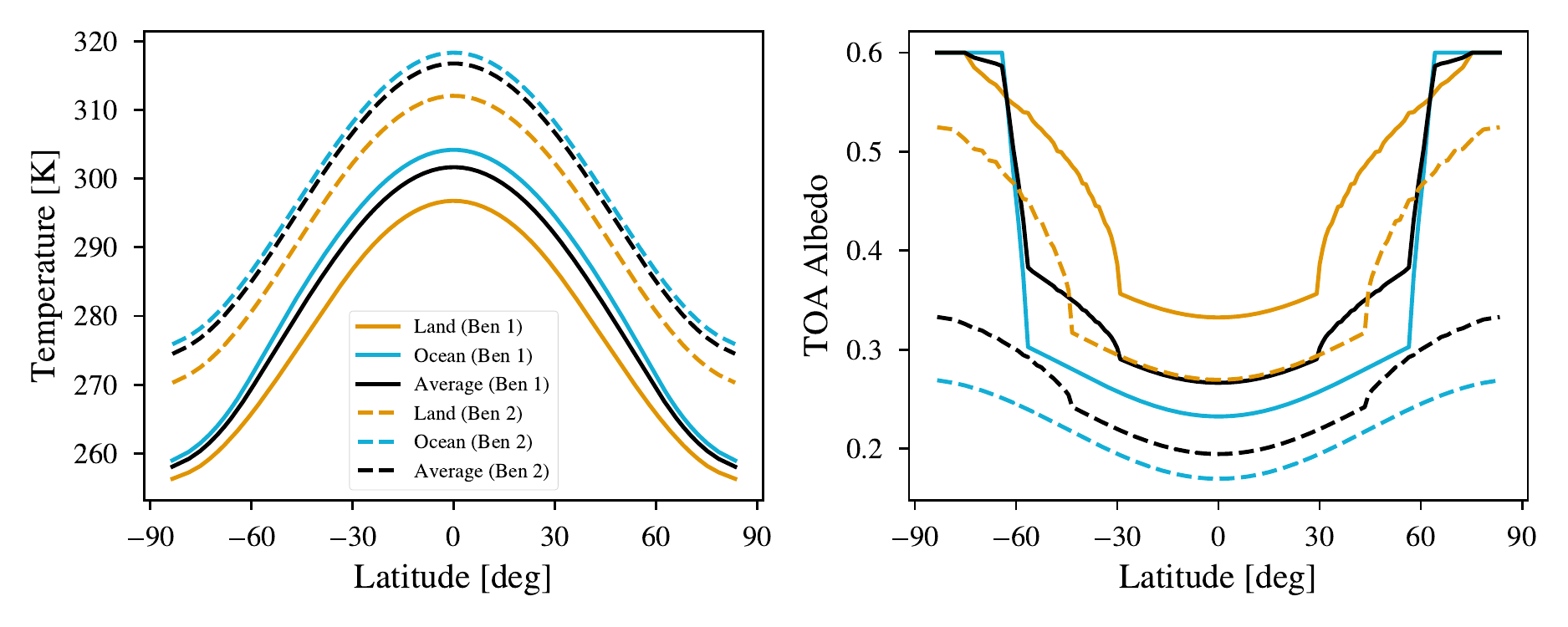}
\caption{Annually-averaged temperature (left) and albedo (right) as a function of latitude for Benchmark 1 and 2 with 75\% ocean and 25\% land at each latitude, using the \texttt{VPLanet/POISE} model. For Benchmark 1, the ice line latitude is $\sim 60^{\circ}$ over the ocean in both hemispheres, indicated by a sharp increase in albedo and temperatures $<$ 271.15 K ($-2^{\circ}$C). Benchmark 2 is substantially warmer, with a mean surface temperature of $\sim 301$ K. The planet in this case has no inter-annual ice. In both cases, ice on land is largely seasonal only, i.e., it melts during summer---this contributes to a higher annually averaged albedo in the mid- to high-latitudes. Note that while \texttt{VPLanet/POISE} explicitly models land and ocean as separate grid cells, this is not a requirement for the FILLET models.}
\label{fig:earth}
\end{figure*}

\begin{figure*}
\centering
\plottwo{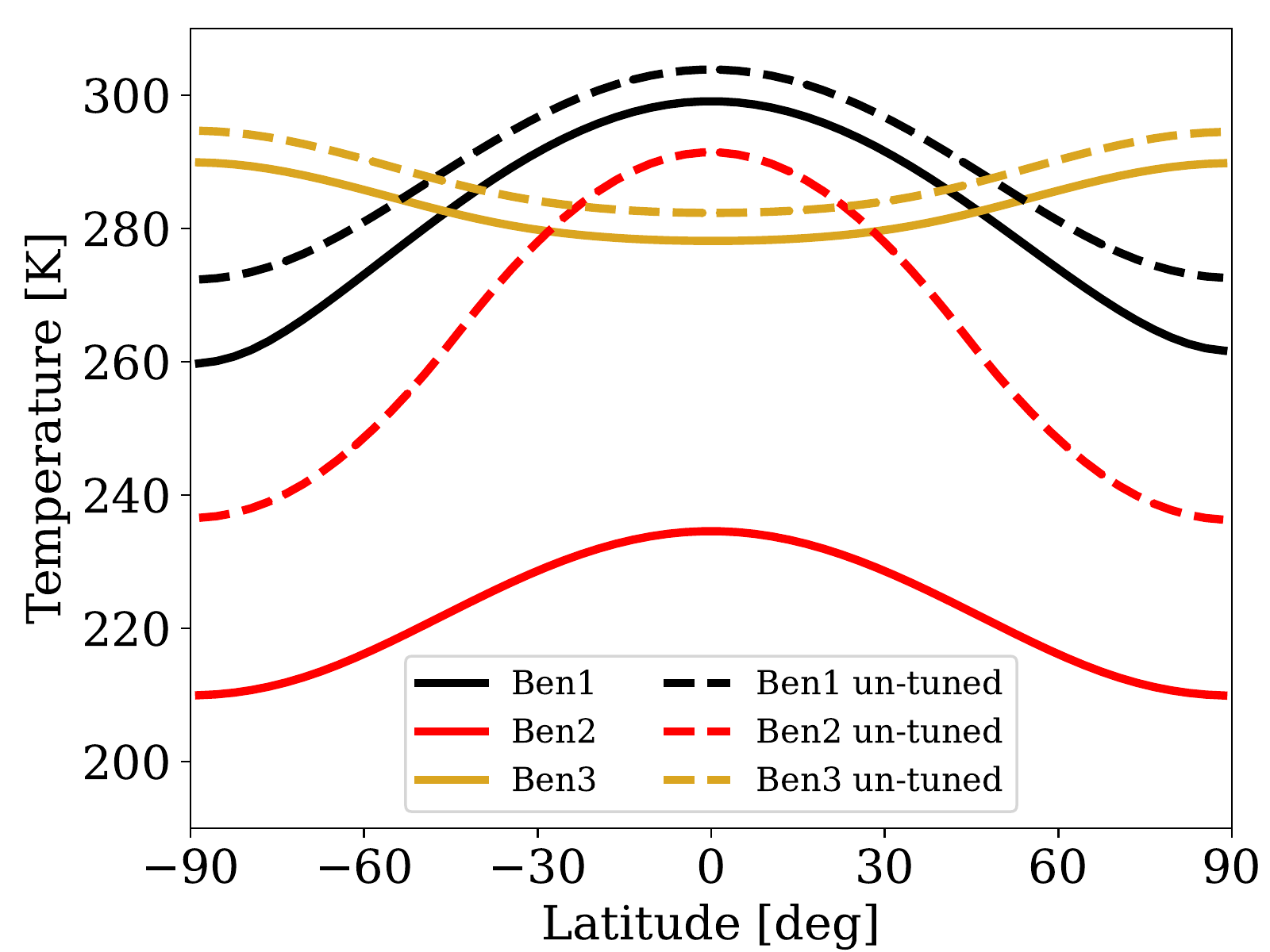}{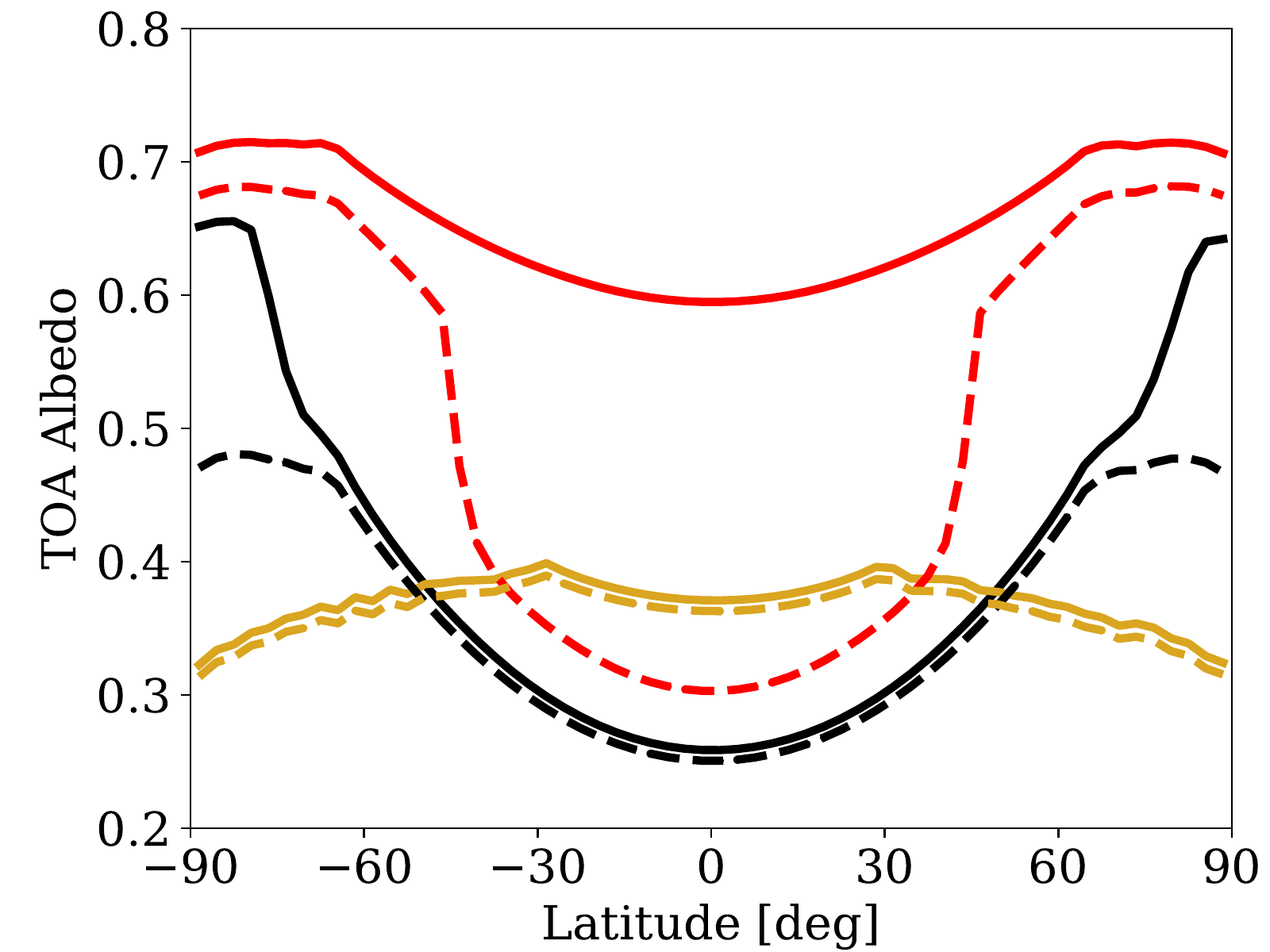}
\caption{Annually-averaged temperature (left) and top-of-atmosphere (TOA) albedo (right) as a function of latitude for Benchmark 1 (black), 2 (red) and 3 (gold) as calculated by \texttt{ESTM}. The value of the parameters unspecified in this paper have been taken either from \citet[dashed]{biasiotti2022} or tuned to produce an average global temperature of 288 K in Benchmark 1 (solid). Tuning has been achieved by lowering the cloud OLR forcing from 26.1 W m$^{-2}$ \citep[the average value for Earth,][]{pierrehumbert2010} to 21.4 W m$^{-2}$. The ice line, when defined as the latitude at which the yearly average surface temperature is below 273.15 K, is $\sim 79^\circ$ and $\sim 61^\circ$ for the un-tuned and tuned Benchmark 1, respectively. For the un-tuned Benchmark 2, it is $\sim 34^\circ$, while the tuned run produces a snowball state. In Benchmark 3 the minimum average temperature is $\sim 278$ K for the tuned and $\sim 282$ K for the un-tuned and the global yearly average ice coverage is $2.7$\% and $1.3$\% respectively. In contrast with \texttt{VPLanet/POISE}, Benchmark 2 runs in \texttt{ESTM} are colder than Benchmark 1 runs. Note that the un-tuned Ben1 and tuned Ben2 and Ben3 runs are not required for FILLET.}
\label{fig:earth_estm}
\end{figure*}

In contrast with Benchmark 1, for Benchmark 2 we specify a set of control parameters common to all models and constrain these to prescribed values. The aim is to compare the models when no tuning is done. For parameters unspecified here, which may be unique to each model, modelers should use defaults or settings from published works. The outgoing longwave radiation, $I$, should be suitable for a 1 bar nitrogen atmosphere with 280 ppm CO$_2$ and water vapor \citep[e.g.,][]{williamskasting97}. Each latitude should be split as 75\% ocean and 25\% land. Models that do not have explicit land/ocean boxes can approximate them by using an averaged albedo and heat capacity. The planet should be given a $23.5^{\circ}$ obliquity and the instellation should be the solar constant, 1361 W m$^{-2}$. Initial conditions for $T$ and $\alpha$ will be each individual model's interpretation of ``warm start'' conditions. Values for input parameters are given in Tables \ref{tab:Params1} and \ref{tab:Params2}. As in Benchmark 1, example output from \texttt{VPLanet/POISE}  is shown in Figure \ref{fig:earth} and output from \texttt{ESTM} is shown in Figure \ref{fig:earth_estm} and Table \ref{tab:estm_ben123}. Figure \ref{fig:earth_estm} and Table \ref{tab:estm_ben123} also show a run of Benchmark 2 using the set of tunings from Benchmark 1---this is not required but may prove useful in identifying the source of model differences. 

\begin{deluxetable}{lccccc}
\tablecaption{Global yearly averages and ice line positions obtained by running \texttt{ESTM} under the conditions specified in the three Benchmarks. \label{tab:estm_ben123}}
\tablecolumns{5}
\tablehead{\colhead{Case} & \colhead{$\langle T \rangle$} & \colhead{$\langle \alpha_\text{TOA}\rangle$} & \colhead{$\langle \text{OLR} \rangle$} & \colhead{Ice fraction} & \colhead{Ice line$^a$} \\
& \colhead{[K]} & & \colhead{[W m$^{-2}$]} & & \colhead{[$^\circ$]}}
\startdata
Benchmark 1 un-tuned & 293.9 & 0.317 & 240.8 & 0.011 & 79 \\
Benchmark 1 tuned & 288.0 & 0.331 & 236.4 & 0.070 & 61 \\
Benchmark 2 un-tuned & 272.7 & 0.426 & 207.1 & 0.349 & 34 \\
Benchmark 2 tuned & 226.5 & 0.633 & 130.2 & 1.000 & 0 \\
Benchmark 3 un-tuned & 285.5 & 0.370 & 226.4 & 0.013 & -$^b$ \\
Benchmark 3 tuned & 281.2 & 0.379 & 223.7 & 0.027 & -$^b$ \\
\enddata
\tablenotetext{a}{Defined as the latitude at which the average temperature is below 273.15 K.}
\tablenotetext{b}{Geographic equator is, on average, colder than the poles. However, no latitudinal band has an average temperature below 273.15 K.}
\end{deluxetable}

\subsection{Benchmark 3: Un-tuned model, high obliquity} 
This benchmark will be identical to Benchmark 2, except with an obliquity of $60^{\circ}$. Here, results may begin to be more sensitive to the parameterizations of $D$, $\alpha$, and $I$. Example output for this case is also given in Figure \ref{fig:earth_estm} and Table \ref{tab:estm_ben123}, for the \texttt{ESTM} model. 

\subsection{Experiment 1: G dwarf warm start}
\begin{figure*}
\centering
\includegraphics[width=0.5\textwidth]{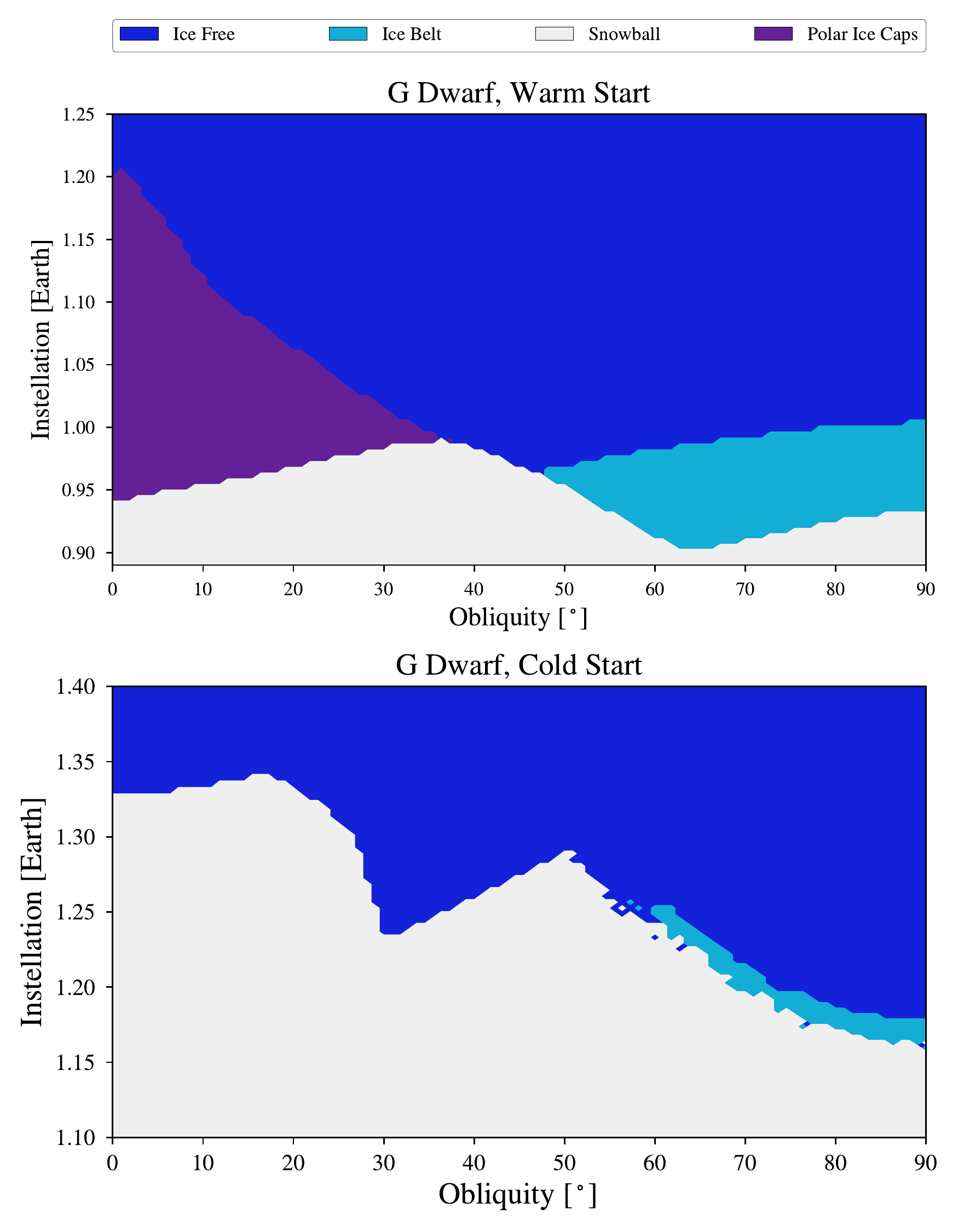}
\caption{Ice state of G dwarf planets as a function of instellation and obliquity. The top panel represents warm start simulations, the bottom  cold start simulations. These simulations are the inspiration for Experiments 1--4. Figure taken from \cite{wilhelm2022}. Settings are as in \citet{wilhelm2022}, many of which differ from those in Tables \ref{tab:Params1} and \ref{tab:Params2}. The code to create this figure is available here: \url{https://github.com/caitlyn-wilhelm/IceCoverage/tree/main/StaticCases/GDwarf}}
\label{fig:Gdwarf}
\end{figure*}

\begin{figure*}
\centering
\includegraphics[width=1.0\textwidth]{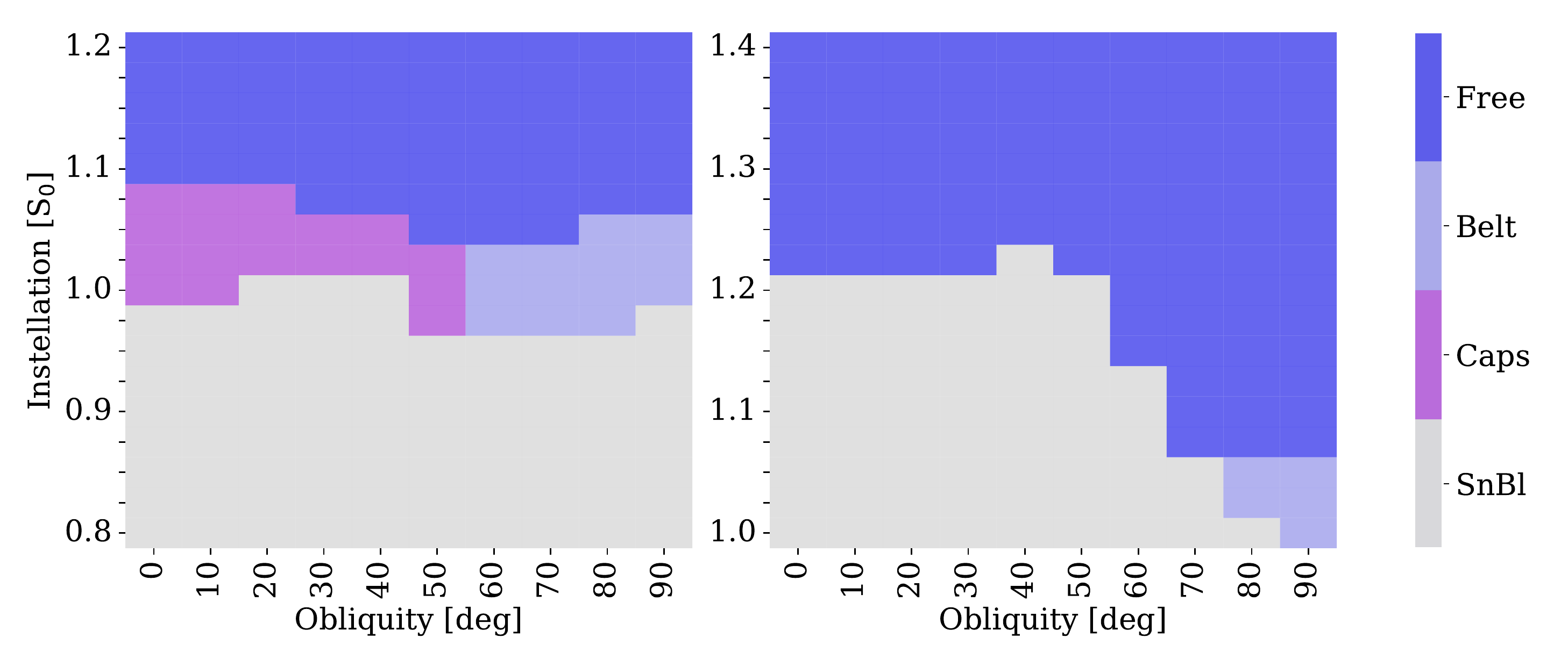}
\caption{Ice states found by \texttt{ESTM} during experiments 1 (left) and 2 (right). The labels in the color bar on the right refer, from top to bottom, to ice free, equatorial ice belt, polar ice caps and Snowball states. Unspecified parameters have been set as in the tuned Benchmark runs. Due to the features of \texttt{ESTM}, the results for $\varepsilon > 60^\circ$ are not reliable.}
\label{fig:experiments12_estm}
\end{figure*}

Here, we create a generalization of Benchmark 1 and 2 to a range of obliquity. We will simulate the same atmosphere as in Benchmarks 2 and 3, with a G dwarf host star, but vary the obliquity from 0$^{\circ}$ to 90$^{\circ}$. Instellation should be varied across a range large enough to encapsulate snowball and ice-free states. The exact values may be model dependent but we propose values in Table \ref{tab:Params1} that have been successful in \texttt{VPLanet/POISE} and \texttt{ESTM}. All simulations should again use a warm start. As with Benchmark 2 and 3, this set of simulations should highlight the uncertainties in $D$, $\alpha$, and $I$. For models that include weathering, we may decide to repeat the same experiment with that feature. Models that include additional ice physics could perform simulations with and without---preliminary testing with \texttt{VPLanet/POISE} indicates that the inclusion of ice sheets makes only a few percent difference in the locations of ice-state instabilities. Figure \ref{fig:Gdwarf} and \ref{fig:experiments12_estm} illustrate the envisioned parameter space for both warm start and cold start (see next Experiment) conditions, using \texttt{VPLanet/POISE} and \texttt{ESTM}, respectively. Figure \ref{fig:Gdwarf} is reproduced from \cite{wilhelm2022}. 

\subsection{Experiment 2: G dwarf cold start}
This experiment is identical to Experiment 1, but models should now be initialized in a ``cold start'', i.e., temperatures below the freezing point of water and corresponding albedos. We should begin to see hysteresis here. Some models may deglaciate at high obliquity. As with Experiment 1, for models that include weathering, we may decide to repeat the experiment done with that feature. Models that include additional ice physics could perform simulations with and without.

\subsection{Experiments 1a and 2a: orbit variations}
We will additionally perform a variation of Experiments 1 and 2 in which semi-major axis should be varied, rather than directly varying the instellation. The semi-major axis ranges are listed in Table \ref{tab:Params1} and are calibrated to cover the same range of instellation as Experiments 1 and 2. Changing semi-major axis has the added effect of changing the year length, which is established as an important parameter in energy balance modeling \citep{rose2017,wilhelm2022}. Settings will otherwise be identical to Experiments 1 and 2. These simulations will still have zero eccentricity.

\subsection{Experiment 3: Bifurcation diagram, varying instellation}
\begin{figure*}
\centering
\includegraphics[width=0.5\textwidth]{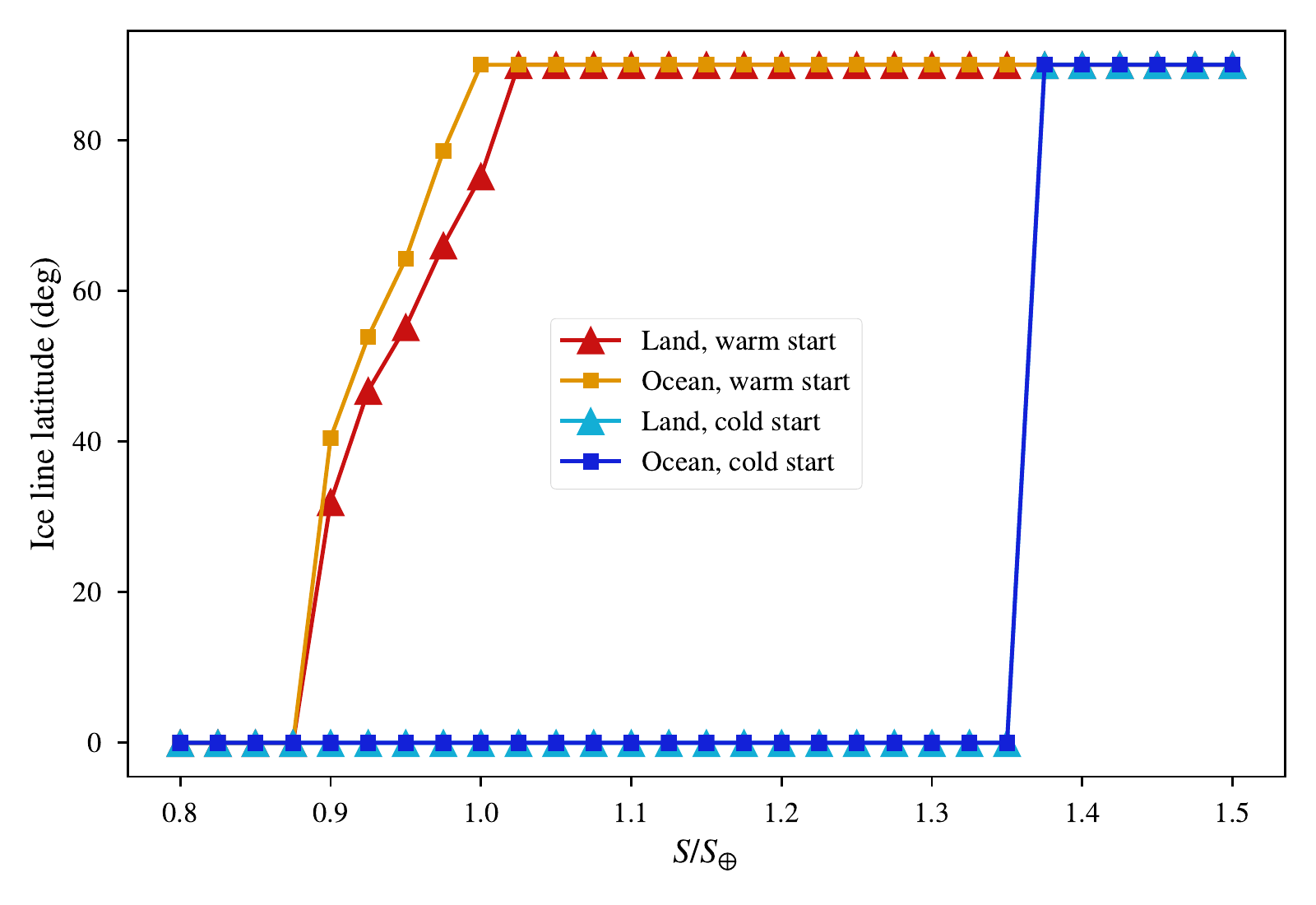}
\caption{Ice line latitude for a G-dwarf planet with Earth's obliquity as a function of instellation. Results are from the \texttt{VPLanet/POISE} model. Ice line latitude $=90^{\circ}$ indicates the ice free state, while ice line $=0^{\circ}$ indicates the snowball state. Individual cases with ice lines between $0^{\circ}$ and $90^{\circ}$ are indicated by squares (ocean) and triangles (land) to make the ``jump'' between states identifiable. In between $S/S_{\oplus}\sim 0.9$ and $\sim1.35$, the climate is bistable, meaning that there are multiple stable states dependent on the initial conditions. Under these settings, cold start conditions never result in an intermediate ice line.}
\label{fig:bifdiag}
\end{figure*}

For this experiment, we will construct bifurcation/hysteresis diagrams. Taking Benchmark 1, we now vary the instellation, starting from both warm and cold starts, until the planet is entirely glaciated or deglaciated, respectively. Here, we will keep CO$_2$ constant at the pre-industrial value. From the output, we will compare the ice line latitude (the boundary between ice-covered and ice-free surfaces). For models with separate land/ocean boxes, there may be two ice lines. There are two potential flavors of this experiment---one in which the orbital period is varied with the instellation (appropriate for a G dwarf host star), and one in which the orbital period is held at 365 days. An example bifurcation diagram is shown in Figure \ref{fig:bifdiag}, using data from \cite{wilhelm2022} and \texttt{VPLanet/POISE}. 

\subsection{Experiment 4: Bifurcation diagram, varying CO$_2$} 
Similar to Experiment 3, we will take Benchmark 1 and construct hysteresis diagrams. This time, however, we will vary the CO$_2$ level in the atmosphere. The behavior is expected to be a strong function of the OLR and albedo parameterizations and so should highlight the differences there. 

\begin{table*}[]
\caption{Protocol parameters (experiment specific). Square brackets indicate spacing of varied parameters.}
\centering
\begin{tabular}{lcccc}
\hline
\hline
 & Instellation & CO$_2$ abundance & Semi-major axis & Obliquity \\
 & $S$ ($S_{\oplus}$)$^\dagger$ & $X_{\text{CO}_2}$ (ppm)$^\ddagger$ & $a$ (au)$^\star$ & $\varepsilon$ ($^{\circ}$) \\
\hline
{\bf Benchmark 2:}\\
Un-tuned, low obliquity & 1 & 280 & 1 & 23.5 \\
\\
{\bf Benchmark 3:}\\
Un-tuned, high obliquity & 1 & 280 & 1 & 60 \\
\\
{\bf Experiment 1:}\\
G-dwarf warm start & 0.8-1.25$^{\star\star}$ [0.025] & 280 & 1 & 0-90 [10$^{\circ}$] \\ 
 \\
{\bf Experiment 1a:}\\
G-dwarf warm start, & Set by $a^{\star\star\star}$ & 280 & 0.875-1.1$^{\star\star}$ [0.0125] & 0-90 [10$^{\circ}$]  \\ 
orbit variation & & &  & \\
\\
{\bf Experiment 2:}\\
G-dwarf cold start & 1.05-1.5$^{\star\star}$ [0.025] & 280 & 1 & 0-90  [10$^{\circ}$] \\
\\
{\bf Experiment 2a:}\\
G-dwarf cold start, & Set by $a^{\star\star\star}$ & 280 & 0.8-0.975$^{\star\star}$ [0.0125]  & 0-90 [10$^{\circ}$]  \\
orbit variation & & & &  \\
\\
{\bf Experiment 3:}\\
Bifurcation diagram,  & 0.8-1.5$^{\star\star}$ [0.0125] & 280 & 1 & 23.5 \\
varying instellation \\
\\
{\bf Experiment 4:}\\
Bifurcation diagram,  & 1 & 50-5050$^{\star\star}$ [100] & 1 & 23.5  \\
varying $X_{\text{CO}_2}$ \\
\\
\hline
\end{tabular}\\
\vspace{0.5em}
$^\dagger$ $S_{\oplus} = 1361$ W m$^{-2}$.\\
$^\ddagger$ CO$_2$ volume mixing ratio in parts-per-million in a 1 bar, N$_2$-dominated atmosphere.\\
$^\star$ 1 au = $1.495978707\times10^{11}$ m.\\
$^{\star\star}$ Ranges should be extended if necessary to capture both snowball and ice-free states.\\
$^{\star\star\star}$ Use the definition $S(a) = S_{\oplus}/a^2$ to scale the instellation. 
\label{tab:Params1}
\end{table*}

\begin{table*}[]
\caption{Protocol parameters (common)}
\centering
\begin{tabular}{cccc}
\hline
\hline
Surface Albedo (land/ocean/ice)  & Heat Diffusion & Heat capacity (land/ocean/ice) & Ocean fraction \\
$\alpha$  & $D$ (W m$^{-2}$ K$^{-1}$) & $C$ (J m$^{-2}$ K$^{-1}$) & \\
\hline
 0.3/0.2/0.6  & 0.5 & $1\times10^7$/$4\times10^8$/$1\times10^7$ & 0.75\\
\\
\hline
\end{tabular}\\
\label{tab:Params2}
\end{table*}

\section{Preliminary analysis}

We offer here a preliminary analysis of the test runs plotted in Figures \ref{fig:earth}-\ref{fig:experiments12_estm}, comparing results from \texttt{VPLanet/POISE} and \texttt{ESTM}. This is not intended to be a thorough analysis but to simply provide a more concrete example of the type of research to follow.

For Ben1 (Figures \ref{fig:earth} and \ref{fig:earth_estm}), \texttt{VPLanet/POISE} and \texttt{ESTM} produce a very similar temperature distribution as a function of latitude (solid black curves in both figures). The temperatures at the equator in both models approach $\sim 300$ K and temperatures at the poles are $\sim 260$ K. While both cases are tuned to produce the same average surface temperature, it is none-the-less encouraging that the equator-to-pole temperature gradient is so similar. The albedos differ qualitatively in shape, though the extremes are similar ($\sim 0.3$ at the equator and $\sim 0.6$ near the poles). The ice line occurs near $61^{\circ}$ latitude in both cases.

For Ben2 (dashed black in Figure \ref{fig:earth} and dashed red in Figure \ref{fig:earth_estm}), we see a marked difference between models. In \texttt{VPLanet/POISE}, Ben2 is $\sim15$ K warmer everywhere than Ben1, while in \texttt{ESTM}, it is $20-35$ K cooler than Ben1, with a steeper equator-to-pole gradient. Here, the albedos differ substantially too. In \texttt{VPLanet/POISE} the albedo stays between 0.2 and 0.35; ice occurs only seasonally in mid- to high-latitudes. In the \texttt{ESTM} case, there are permanent ice caps that are much larger than in Ben2---the ice line extends to $\sim 34^{\circ}$ latitude.

The key difference between the two models is most likely the treatment of radiation: \texttt{VPLanet/POISE} uses the simply linearized formula for the OLR (Section \ref{sec:protocol}) and a simple parameterization for the albedo, while \texttt{ESTM} uses the lookup-tables derived from the radiative-convective model \texttt{EOS} for both the OLR and albedo. Thus the use of the same surface properties (Table \ref{tab:poiseinput}) results in a substantially different climate state in the two models. The simple albedo formulation in \texttt{VPLanet/POISE} results in sharp jumps in the TOA albedo when ice is present, while the more complex treatment in \texttt{ESTM} smoothes out these jumps.

For Experiments 1 and 2, \texttt{VPLanet/POISE} and \texttt{ESTM} produce the same set of final climate states, though the boundaries of each vary substantially (Figure \ref{fig:Gdwarf} and \ref{fig:experiments12_estm}). In the warm start cases, the snowball state generally extends to higher instellation for \texttt{ESTM} than for \texttt{VPLanet/POISE}. This dovetails with the results of Ben2, which found \texttt{ESTM} to be colder. The polar caps have less extent in both directions in instellation for \texttt{ESTM}, but extends to higher obliquity ($\sim 55^{\circ}$, compared to $\sim 35^{\circ}$ for \texttt{VPLanet/POISE}). Ice belts systematically occur at a lower instellation for \texttt{VPLanet/POISE} and extend to lower obliquity, compared to \texttt{ESTM}.

In both models, polar caps are non-existent for cold start conditions. Ice belts occur in a tiny region above $\sim 75^{\circ}$ obliquity for \texttt{ESTM} and a small finger along the snowball state above $\sim 60^{\circ}$ obliquity for \texttt{VPLanet/POISE}. In this case, at high obliquity, \texttt{ESTM} is systematically warmer than \texttt{VPLanet/POISE}, contrary to the warm start cases, where \texttt{ESTM} is colder. This may speak to the inclusion of an explicit ice sheet model in \texttt{VPLanet/POISE}, which increases the thermal inertia and makes it harder for the planet to escape the snowball state. Interestingly, the low obliquity snowball ``hump'' seen in \texttt{VPLanet/POISE} (Figure \ref{fig:Gdwarf}, lower panel) doesn't appear at all in \texttt{ESTM}. More detailed analysis will be required to isolate the cause of this feature.

\section{Standard output for comparison}

EBM output is generally small enough that we can use plain text/ASCII output. For each simulation, there should be an output file containing latitudinally-varying outputs, the columns of which are listed in the top portion of Table \ref{tab:Outputs}. An additional output file should be generated for each Experiment, containing the global information needed to make contour and hysteresis plots. The columns are listed in the lower portion of Table \ref{tab:Outputs}. Each output parameter is given a short name that should be used for the column headers in the output. This will facilitate comparison by making it easy for analysis code to find the correct quantities. For models that are unable to use the simplified constants in Table \ref{tab:Params2}, additional columns should be output that contain the heat diffusion, heat capacity, and ocean fraction, as necessary. One important thing to note is that each model should provide a clear definition of the ice line latitude. Some examples are: where the annually averaged temperature falls below freezing; the average latitude where the temperature falls below freezing; etc. Latitudinally-varying quantities used for the comparison should be annually-averaged, though modelers should strive to retain seasonally-varying quantities as these values may be useful to diagnose inter-model differences.

Up-to-date information regarding FILLET is located at \url{https://ckan.emac.gsfc.nasa.gov/organization/cuisines-fillet}. Inputs described in this protocol and scripts related to the analysis of data and production of plots for the publications will be made available on the FILLET GitHub repository at \url{https://github.com/projectcuisines/fillet}.  Inputs will be available immediately while scripts to reproduce results will be made publicly available upon the publication of the results. %Make sure to actually add these!

\begin{table*}[]
\caption{Outputs}
\centering
\begin{tabular}{lcc}
\hline
\hline
Output type & Columns (units) & Output header name\\
\hline
{\em Latitudinally-varying outputs:}
 & Latitude ($^{\circ}$) & Lat\\
 & Annually-averaged surface temperature (K) & Tsurf\\
 & Annually-averaged surface albedo & Asurf\\
 & Annually-averaged TOA albedo & ATOA\\
\\
{\em Global outputs:}
 & Instellation ($S_{\oplus}$)$^{\dagger}$ & Inst\\
 & Obliquity ($^{\circ}$) & Obl\\
 & $X_{\text{CO}_2}$ (ppm)$^\ddagger$ & XCO2 \\
 & Global mean surface temperature (annual average) (K) & Tglob\\
 & North ice line latitude (land and ocean or average) ($^{\circ}$) & IceLineN\\
 & South ice line latitude (land and ocean or average) ($^{\circ}$) & IceLineS\\
\hline
\end{tabular}\\
$^{\dagger}$ $S_{\oplus} = 1361$ W m$^{-2}$.\\
$^\ddagger$ CO$_2$ volume mixing ratio in parts-per-million \\
\label{tab:Outputs}
\end{table*}

\section{Summary}
\label{sec:summary}
Here, we have detailed the protocol for the FILLET model intercomparison project, one aspect of the CUISINES framework. Our aim is to determine the most critical components of EBM modeling of exoplanets and to provide a standard to which future EBM users or developers can compare. This work will promote consistency across exoplanet EBMs and provide a nuanced understanding of the strengths and weaknesses of each model. To that end, we have designed a set of experiments that will benchmark the most basic ingredients of the EBM. These numerical experiments are low barrier-of-entry and should be within the reach of most EBMs. We invite participation from all members of the community with such a model. Assuming that there is sufficient interest within the community, the exoMIP will continue beyond the experiments outlined here and will expand to include additional model components, such as carbon cycling and different surface models. 

%\textbf{This work describes the first phase of the FILLET project, which may or may not entail additional phases, depending on the outcomes of the first. Some potential avenues for follow-up phases have been discussed, such as the effects of orbital eccentricity, carbon cycling, slow or synchronous rotation, Milankovitch cycles, and combinations of all of the above. However, for now the contours of these additional phases are not yet determined. }

\appendix

\section{Model input parameters}

For completeness, we compile a list of input parameters for the currently participating models. This is intended to facilitate the comparison between models and assist with the model set up. Input parameters for each model are given in Table \ref{tab:poiseinput} (\texttt{VPLanet/POISE}), Table \ref{tab:hextorinput} (\texttt{HEXTOR}), Table \ref{tab:ram1dinput} (\texttt{OPS} and \texttt{PlaHab}), Table \ref{tab:ktinput} (Kadoya-Tajika model), and Table \ref{tab:estminput} (\texttt{ESTM}) . 

\begin{deluxetable*}{lcccc}
\tabletypesize{\scriptsize}
\tablewidth{0pt}
\tablecaption{\texttt{VPLanet/POISE} input parameters (seasonal model only---the annual model has many independent settings but is not used in this exoMIP) \label{tab:poiseinput}}
\tablehead{ 
\colhead{Category} & \colhead{Symbol}  & \colhead{Description (units)} & \colhead{Name in model}& \colhead{Notes}
} 
\startdata
{\bf General} & & & & \\
& - & Number of latitude cells & iLatCellNum & Cells are equal area, not width \\
& - & Type of land distribution & sGeography & Used for preset land fractions \\
& $f_{\mathrm{land}}$ & Land fraction & dLandFrac & Only used when sGeography = uniform \\
& - & Number of time steps in year & iNStepInYear & ``Year'' = orbital period \\
& - & Number of years to run EBM & iNumYears & Use for coupling to ice sheet, orbit models\\
& $P_{\mathrm{rot}}$  & Rotation period (days) & dRotPeriod & - \\
\hline
{\bf Heat} & & & & \\
{\bf ~~capacity}& $C_{\mathrm{land}}$ & Land heat capacity (J m$^{-2}$ K$^{-1}$) & dHeatCapLand & - \\
& $C_{\mathrm{ocean}}$ & Ocean heat capacity (J m$^{-3}$ K$^{-1}$) & dHeatCapWater & Multiplied by $m_d$ to get areal heat capacity \\
& $m_d$ & Ocean mixing depth (m) & dMixingDepth & Used in areal heat capacity of ocean \\
\hline
{\bf Heat} & & & & \\
{\bf ~~diffusion}& $D$ & Latitudinal diffusion coefficient (W m$^{-2}$ K$^{-1}$) & dDiffusion & Constant with latitude, typically \\
& $\nu$ & Land-ocean diffusion coefficient (W m$^{-2}$ K$^{-1}$) & dNuLandWater & Mislabeled as unitless in \citet{deitrick2018} \\
& - & Increase $D$ to mimic Hadley cell & bHadley & Only functions at low latitudes \\
& - & Set $D = B/4$ & bMEPDiff & Based on maximum entropy production \\
& & & & \citep{lorenz2001} \\
& - & Adjusts $D$ for rotation rate & bDiffRot & Based on \citet{williamskasting97} \\
\hline
{\bf OLR} & & & & \\
& A & OLR constant (W m$^{-2}$) & dPlanckA & From approximation $I \approx A + BT$\\
& B & OLR linear coefficient (W m$^{-2}$ K$^{-1}$) & dPlanckB & " \\
& - & Calculate $(A,B)$ as a function of $(T,p_{\mathrm{CO}_2})$ & bCalcAB & Overrides dPlanckA and dPlanckB \\
& $p_{\mathrm{CO}_2}$ & Partial pressure of CO$_2$ & dpCO2 & Use with bCalcAB \\
& - & Set which OLR parameterization is used & iOLRModel & " \\
\hline
{\bf Instellation} & & & & \\
 & $L_{\star}$ & Luminosity of host star (W) & dLuminosity &  Set in host star input file\\
 & $R_{\star}$ & Radius of host star (au or m) & dRadius & Set in host star input file\\
 & $a$ & Semi-major axis of planet (au) & dSemi & -\\
 & $e$ & Orbital eccentricity & dEcc & -\\
 & $\varepsilon$ & Obliquity (deg) & dObl & -\\
 & $\psi$ & Precession parameter/angle of equinox (deg) & dPrecA & -\\
 & $\varpi$ & Longitude of pericenter of orbit (deg) & dLongP & Defined as $\varpi = \omega + \Omega$\\
 & $\omega$ & Argument of pericenter of orbit (deg) & dArgP & User can set dLongP or both dArgP \\
 & $\Omega$ & Longitude of ascending node of orbit (deg) & dLongA & and dLongA, but not all three\\
\hline
{\bf Albedo} & & & & \\
& $\alpha_{\mathrm{ice}}$ & Surface albedo of ice/snow & dIceAlbedo & - \\
& $\alpha_{\mathrm{land}}$ & Surface albedo of land & dAlbedoLand & Usually tuned to include clouds \\
& $\alpha_{\mathrm{ocean}}$ & Surface albedo of ocean & dAlbedoWater & " \\
& - & Freezing temperature of ocean ($^{\circ}$C) & dFrzTSeaIce & Controls transition to ice albedo \\
\hline
{\bf Initial} & & & & \\
{\bf ~~conditions}& $T_{\mathrm{init}}$ & Initial estimate of mean temperature ($^{\circ}$C) & dTGlobalInit & $T$ is set via $T = (T_{\mathrm{init}}-7.35)+20(1-2x^2)$, \\
& & & & where $x=\sin{(\mathrm{latitude})}$ \\
& - & Use cold start conditions & bColdStart & Subtracts $40^{\circ}$C from above equation\\
\hline
{\bf Sea ice} & & & & \\
{\bf ~~model}& - & Use sea ice model (thickness and insulation) & bSeaIceModel & Slows computation substantially\\
& - & Sea ice heat conductivity (W m$^{-1}$ K$^{-1}$) & dSeaIceConduct & - \\
\hline
{\bf Ice sheet} & & & & \\
{\bf ~~model}& - & Couple ice sheet model (land) to EBM & bIceSheets &- \\
& - & Initial height of ice sheet (m)  & dInitIceHeight &- \\
& - & Initial ice sheet latitude (deg) & dInitIceLat & Assumes polar ice cap\\
& - & Time step of ice sheet model (orbits) & iIceDt & -\\
& - & Re-run Seasonal EBM after \# orbits & iReRunSeas &-\\
& - & Fudge factor in albation equation & dAblateFF & Used to tune ice melt rate\\
& - & Minimum ice height (m) & dMinIceSheetHeight & Below this value, ice is set to zero\\
& $r_{\mathrm{snow}}$ & Deposition rate of ice/snow (kg m$^{-2}$ s$^{-1}$) & dIceDepRate &- \\
\enddata
\end{deluxetable*}

\begin{deluxetable*}{lcccc}
\tabletypesize{\scriptsize}
\tablewidth{0pt}
\tablecaption{\texttt{HEXTOR} input parameters\label{tab:hextorinput}}
\tablehead{ 
\colhead{Category} & \colhead{Symbol}  & \colhead{Description (units)} & \colhead{Name in model}& \colhead{Notes}
} 
\startdata
{\bf General} & & & & \\
& - & Use seasonal model & seasonflag & Annual model used if false \\
& - & Total calculation length (s) & tend & - \\
& - & Time step size (s) & dt & - \\
& - & Rotation rate (rad/s) & rot & - \\
& $p_{\mathrm{surf}}$ & Surface pressure (bar) & pg0 & - \\
& - & Ocean coverage (percent) & ocean & - \\
& - & Set land/ocean configuration & igeog & - \\
& - & Number of years per time step & yrstep & Only affects output labeling\\
& $M_{\star}$ & Mass of host star (g) & msun & Used for orbital calculation \\
& - & Enable synchronous rotation mode & do\_longitudinal & -\\
\hline
{\bf Heat} & & & & \\
{\bf ~~capacity}& $C_{\mathrm{land}}$ & Land heat capacity (J m$^{-2}$ K$^{-1}$) & heatcap & - \\
& - & Use constant heat capacity at all grid points & constheatcap & - \\
\hline
{\bf Heat} & & & & \\
{\bf ~~diffusion}& $D$ & Thermal diffusion coefficient (W m$^{-2}$ K$^{-1}$) & d0 & - \\
& - & Adjust diffusion coefficient & diffadj & Uses scaling from  \citet{williamskasting97} \\ 
\hline
{\bf OLR} & & & & \\
& - & Radiation parameterization & radparam & - \\
& $X_{\mathrm{CO}_2}$ & CO$_2$ mixing ratio & fco2 & - \\
& $X_{\mathrm{H}_2}$ & H$_2$ mixing ratio & fh2 & \citet{hayworth2020} parameterization only\\
& - & Reduction in OLR due to clouds (W m$^{-2}$) & cloudir & - \\
& - & Set OLR to linear form & linrad & Use $I = A + BT$ \\
\hline
{\bf Instellation} & & & & \\
& - & Relative solar constant & relsolcon & - \\
& $S_0$ & Stellar/solar constant (W m$^{-2}$) & solarcon & - \\
& - & Read solar constant from file & soladj & - \\
& - & Read orbital parameters from file & do\_manual\_seasons & - \\
& $a$ & Semi-major axis (cm) & a & - \\
& $e$ & Orbital eccentricity & ecc & - \\
& $\varepsilon$ & Obliquity (deg) & obl & - \\
& $\varpi$ & Longitude of perihelion (deg) & peri & Measured w.r.t. vernal equinox\\
\hline
{\bf Albedo} & & & & \\
& $\alpha_{\mathrm{ice}}$ & Surface albedo of snow & snowalb & - \\
& $\alpha_{\mathrm{land}}$ & Land surface albedo & groundalb & - \\
& $\alpha_{\mathrm{ocean}}$ & Ocean surface albedo & ocnalb & - \\
& - & Percent of land in each grid point with snow/ice & landsnowfrac & - \\ 
& $f_{\mathrm{cloud}}$ & Fractional cloud cover & fcloud & Modifies surface albedo \\
& - & Reduce albedo by factor $f_{\mathrm{cloud}}$ & cloudalb & - \\
& - & Set surface albedo to a constant & linalb & - \\
\hline
{\bf Initial} & & & & \\
{\bf ~~conditions}& $T_{\mathrm{init}}$ & Initial temperature at all grid points (K) & tempinit & \\
\hline
{\bf Carbonate-} & & & & \\
{\bf silicate cycle} & - & Enable carbonate-silicate cycle & do\_cs\_cycle & Adjusts $X_{\mathrm{CO}_2}$\\
& $V$ &  Volcanic outgassing rate (bar Gyr$^{-1}$) & outgassing & - \\
& $W$ & Weathering rate (bar Gyr$^{-1}$) & weathering & - \\
& $\beta$ & Beta exponent for weathering  & betaexp & - \\
& $k_{\mathrm{act}}$ & Activation energy for weathering (K$^{-1}$) & kact & - \\
& $k_{\mathrm{run}}$ & Runoff factor for weathering (K$^{-1}$) & krun & - \\
& - & Enable H$_2$ adjustment from outgassing/escape & do\_h2\_cycle & \citet{hayworth2020} parameterization only \\
& - & Hydrogen outgassing rate (Tmol yr$^{-1}$) & h2outgas & - \\
\hline
{\bf Stochasticity} & & & & \\
& - & Add random noise to EBM equation & do\_stochastic & - \\
& - & Magnitude of noise variable &noisevar & - \\
\enddata
\end{deluxetable*}

\begin{deluxetable*}{lcccc}
\tabletypesize{\scriptsize}
\tablewidth{0pt}
\tablecaption{\texttt{OPS} and \texttt{PlaHab} input parameters \label{tab:ram1dinput}}
\tablehead{ 
\colhead{Category} & \colhead{Symbol}  & \colhead{Description (units)} & \colhead{Name in model}& \colhead{Notes}
} 
\startdata
{\bf General} & & & & \\
& - & Use seasonal model & seasonflag & Annual model used if false \\
& - & Total calculation length (s) & tend & - \\
& - & Convergence criterion (K) & cnvg & Interannual $\Delta T$ threshold\\
& $P_{\mathrm{rot}}$ & Rotation period (s) & lday & - \\
& $p_{\mathrm{dry}}$ & Surface pressure without H$_2$O (bar) & Pdry & - \\
& $g$ & Gravity (m s$^{-2}$) & g & - \\
& - & Ocean coverage (percent) & ocean & - \\
& - & Set land/ocean configuration & igeog & - \\
& $M_{\star}$ & Mass of host star (M$_{\odot}$) & smass & {Influences the orbital period.} \\
& - & Number of latitudes & nbelts & - \\
& - & Number of longitudes & longbelts &  \texttt{PlaHab} only\\
\hline
{\bf Heat} & & & & \\
{\bf ~~capacity}& $C_{\mathrm{land}}$ & Land heat capacity (J m$^{-2}$ K$^{-1}$) & heatcap & - \\
\hline
{\bf Heat} & & & & \\
{\bf ~~diffusion}& $D$ & Latitudinal thermal diffusion coefficient (W m$^{-2}$ K$^{-1}$) & d0 & Adjusted for rotation, pressure, molecular mass \\
&$D_{\mathrm{long}}$ & Longitudinal hermal diffusion coefficient (W m$^{-2}$ K$^{-1}$) & d0l & Same dependence as $D$, \texttt{PlaHab} only \\
\hline
{\bf OLR} & & & & \\
& $p_{\mathrm{CO}_2}$ & Initial CO$_2$ pressure & pco2i & {Updated with carbon cycle.}\\
& $L$ & Latent heat of vaporization of water (J kg$^{-1}$) & L & Used for cloud IR param. \\
& $L_{\mathrm{CO}_2}$ & Latent heat of vaporization of CO$_2$ (J kg$^{-1}$) & LCO2 & " \\
& $C_d$ & Surface drag coefficient & Cd & " \\
& $v$ & Near surface wind speed (m s$^{-1}$) & vel & " \\
& - & Surface convective heat flux (W m$^{-2}$) & FE &  " \\
\hline
{\bf Instellation} & & & & \\
& - & Spectral type of host star & STAR & - \\
& $a$ & Semi-major axis (au) & a0 & - \\
& $e$ & Orbital eccentricity & ecc & - \\
& $\varepsilon$ & Obliquity (deg) & obl & - \\
& $\varpi$ & Longitude of perihelion (deg) & peri & Measured w.r.t. vernal equinox\\
\hline
{\bf Albedo} & & & & \\
& $\alpha_{\mathrm{land}}$ & Land surface albedo & groundalb & - \\
& - & Percent of land in each grid point with snow/ice & landsnowfrac & - \\ 
\hline
{\bf Initial} & & & & \\
{\bf ~~conditions} & - & Start with snowball conditions & snowballflag  & $T_{\mathrm{surf}} = 190$ K\\
& $T_{\mathrm{init}}$ & Initial surface temperature (K) & tempi & Applied to all latitudes\\
\enddata
\end{deluxetable*}

\begin{deluxetable*}{lcccc}
\tabletypesize{\scriptsize}
\tablewidth{0pt}
\tablecaption{Kadoya \& Tajika model input parameters  \label{tab:ktinput}}
\tablehead{ 
\colhead{Category} & \colhead{Symbol}  & \colhead{Description (units)} & \colhead{Name in model}& \colhead{Notes}
} 
\startdata
{\bf General} & & & & \\
& - & Number of latitude cells & nLat & - \\
& $f_{\text{ocean}}$ & Ocean fraction in each cell & f\_ocean & 
    Uniform in all latitudinal grids
\\
& $P_0$ & Atmospheric surface pressure (bar) & P0 & $P_\text{0} = 1 + p_{\mathrm{CO}_2} + p_{\mathrm{H}_2\mathrm{O}}$ \\
& $p_{\mathrm{H}_2\mathrm{O}}$ & Partial pressure of water (bar) & pH2O & Calculated with a globally averaged temperature \\
& - & Number of time steps per orbit & nPeri & - \\
\hline
{\bf Heat} & & & & \\
{\bf ~~capacity}& $C_{\mathrm{land}}$ & Land heat capacity (J m$^{-2}$ K$^{-1}$) & C\_land & - \\
& $C_{\mathrm{ocean}}$ & Ocean heat capacity (J m$^{-2}$ K$^{-1}$) K$^{-1}$) & C\_ocean & - \\
& $C_{\mathrm{ice}}$ & Heat capacity over ice (J m$^{-2}$ K$^{-1}$) K$^{-1}$) & C\_ice & - \\
\hline
{\bf Heat} & & & & \\
{\bf ~~diffusion}& $D_0$ & Diffusion coefficient of 1 bar atmos. (W m$^{-2}$ K$^{-1}$) & D0 & Scales with surface pressure \\
\hline
{\bf OLR} & & & & \\
& $p_{\mathrm{CO}_2}$ & Partial pressure of CO$_2$ (bar) & pCO2 & -\\
& - & Offset in OLR due to clouds (W m$^{-2}$) & IRcloud & - \\
\hline
{\bf Instellation} & & & &\\
& $S_\text{eff}$ & Effective solar constant & Seff & Multiples of $S_\text{earth}$; $S_\text{earth}=1366\ \text{W m}^{-2}$ \\
& $a$ & Semi-major axis (au)& sma & - \\
& $e$ & Eccentricity of orbit & ecc &  - \\
& $\varepsilon$ & Obliquity (deg) & obl & -\\
& $\varpi$ & Angle between pericenter and equinox (deg) & ps & -\\
\hline
{\bf Albedo} & & & &\\
& $\alpha_{\mathrm{ice}}$ & Surface albedo of ice & a\_ice & - \\
& $\alpha_{\mathrm{land}}$ & Surface albedo of land & a\_land &- \\
& $\alpha_{\mathrm{ocean}}$ & Surface albedo of ocean & a\_ocean & - \\
& $\alpha_{\mathrm{cloud}}$ & Albedo of clouds & a\_cloud & Depends on zenith angle \\
& $f_{\mathrm{cloud}}$ & Cloud coverage & f\_cloud & - \\
& $f_{\mathrm{ice}}$ & Sea ice coverage & f\_ice & See Eq. 11 in \citet{kadoya2019}\\
\enddata
\end{deluxetable*}

\clearpage

\startlongtable
\begin{deluxetable*}{lcccc}
\tabletypesize{\scriptsize}
\tablewidth{0pt}
\tablecaption{\texttt{ESTM} input parameters \label{tab:estminput}}
\tablehead{ 
\colhead{Category} & \colhead{Symbol}  & \colhead{Description (units)} & \colhead{Name in model}& \colhead{Notes}
} 
\startdata
{\bf General} & & & & \\
& - & Number of latitude cells                 & N         & - \\
& - & Number of time steps in one orbit        & Ipoints   & - \\
& $M_{\star}$ & Mass of the parent star (M$_\Sun$)       & MStar     & Used to calculate the orbital period \\
& - & Land distribution                        & gg        & Preset distributions of lands \\
& $f_{\mathrm{ocean}}$ & Ocean fraction                           & fo\_const & Only used if gg=0 \\
\hline
{\bf Heat} & & & & \\
{\bf ~~capacity} & $C_{\mathrm{land}}$ & Land heat capacity (J m$^{-2}$ K$^{-1}$)        & CL  & - \\
& $C_{\mathrm{ocean}}$ & Ocean heat capacity (J m$^{-2}$ K$^{-1}$)             & CO  & - \\
& $C_{\mathrm{ice,l}}$  & Heat capacity of ice over land (J m$^{-2}$ K$^{-1}$)  & CIL & - \\
& $C_{\mathrm{ice,o}}$   & Heat capacity of ice over ocean (J m$^{-2}$ K$^{-1}$) & CIO & - \\
\hline
{\bf Heat} & & & & \\
{\bf ~~diffusion}& $D_0$ & Latitudinal diffusion normalization (W m$^{-2}$ K$^{-1}$) & D0par & $D=\text{D0par} \times f(\text{Rplan},\text{Prot}...)$ \\
& - & Planetary radius (R$_\Earth$)                                               & Rplan    & - \\
& $P_{\mathrm{rot}}$  & Planetary rotation period (days)                                            & Prot     & - \\
& $p_{\mathrm{dry}}$ & Surface dry (no H$_2$O) pressure (Pa)                & pressdry & Usually taken from the RT lookup tables \\
& $C_p$ & Dry heat capacity; isobaric (J kg$^{-1}$ K$^{-1}$) & cp       & Usually taken from the RT lookup tables \\
& - & Dry mean molecular weight (g mol$^{-1}$)                   & molwt    & Usually taken from the RT lookup tables \\
& $g$ & Gravity acceleration at surface (m s$^{-1}$)                                & g        & Usually taken from the RT lookup tables \\
& - & Relative humidity (unit fraction)                                                       & rh       & Usually taken from the RT lookup tables \\
& - & Modulation of latitudinal diffusion                                         & R        & Enhances meridional diffusion at low\\
& & & & zenith angle (i.e. in the tropical region)\\
& - & Ratio of moist over dry eddy transport                                      & lambda0  & - \\
& - & Exponent of the Rplan scaling of $D$                  & Rplan\_expo & - \\
& - & Exponent of the Prot scaling of $D$                  & Vrot\_expo  & - \\
\hline
{\bf OLR} & & & & \\
& - & \multicolumn{3}{c}{OLR and TOA albedo are provided in the form of lookup tables} \\
\hline
{\bf Instellation} & & & &\\
& $L_\star$ & Luminosity of the host star (L$_\Sun$)                       & LumStar & - \\
& -         & Fraction of LumStar actually used in the run                 & LumEvol & Added for evolutionary studies \\
& $a$       & Semi-major axis (au)                                         & smaP    & - \\
& $e$       & Eccentricity of orbit                                        & eccP    & - \\
& $\varepsilon$ & Obliquity (deg)                                          & obliq   & - \\
& $\omega$  & Argument of pericenter of orbit (deg)                        & omegaPERI & - \\
\hline
{\bf Albedo} & & & &\\
& $\alpha_{\mathrm{land}}$  & Surface albedo of land                      & asl   & Value given for zenith angle $=60^\circ$ \\
&  $\alpha_{\mathrm{ice,l}}$                        & Surface albedo of ice over land             & asils & Value given for zenith angle $=60^\circ$ \\
& $\alpha_{\mathrm{ocean}}$ & Surface albedo of ocean                     & aso   & See \citet{briegleb1986} \\
& $\alpha_{\mathrm{ice,o}}$     & Surface albedo of ice over ocean            & asils & Value given for zenith angle $=60^\circ$ \\
& -                         & Coefficient of the land albedo zenith-dependence & d & See \citet{briegleb1986,briegleb1992} \\
\hline
{\bf Initial conditions} & & & &\\
& $T_{\mathrm{init}}$ & Starting temperature (K)  & Tstart     & - \\
& - & Initial mean orbital anomaly & Ls0\_spring & - \\
\hline
{\bf Ice fraction} & & & & \\
{\bf ~~coverage} & - & Turning point temperature of ice fraction over land & T0il & Generalized logistic function parameter \\
& -               & Growth rate of ice fraction over land                & GRil & Generalized logistic function parameter \\
& -               & Shape parameter of ice fraction over land            & SPil & Generalized logistic function parameter \\
& -               & Turning point temperature of ice fraction over ocean & T0io & Generalized logistic function parameter \\
& -               & Growth rate of ice fraction over ocean               & GRio & Generalized logistic function parameter \\
& -               & Shape parameter of ice fraction over ocean           & SPio & Generalized logistic function parameter \\
\hline
{\bf Cloud} & & & & \\
{\bf ~~model} & $f_{\mathrm{cloud,l}}$ & Cloud coverage over land  & fcl & - \\
& $f_{\mathrm{cloud,o}}$        & Cloud coverage over ocean & fco & - \\
& $f_{\mathrm{cloud,i}}$       & Cloud coverage over ice   & fci & Limiting value for 0\% global ice cover \\
& $f_{\mathrm{cloud,i}}$          & Cloud coverage over ice when in snowball             & fci\_snowball & Limiting value for 100\% global ice cover \\
& $\alpha_{\mathrm{cloud}}$  & Albedo of clouds                                          & ac0      & - \\
& -           & Slope of linear dependence between  & acm      & $z=$ zenith angle; \\
& &ac0 and $\mu=\cos(z)$  & & see \citet{williamskasting97} \\
& -           & Cloud shortwave transmittance                             & ct0      & Asymptotic value at high global mean $T$ \\
& -           & Cloud shortwave transmittance in snowball        & ct\_snowball & Asymptotic value at low global mean $T$ \\
& -           & Mid-point temperature of cloud    & tTemp0   & Dependence modeled as tanh \\
&     & shortwave transmittance    &  &  \\
& -           & Slope control of cloud shortwave transmittance            & tTempDiv & Dependence modeled as tanh \\
& -           & Cloud OLR forcing (W m$^{-2}$)                            & CRE0     & Asymptotic value at high global mean $T$ \\
& -           & Cloud OLR forcing in Snowball (W m$^{-2}$)           & CRE\_snowball & Asymptotic value at low global mean $T$ \\
& -           & Mid-point temperature of cloud OLR forcing                & CRETemp0 & Dependence modeled as tanh \\
& -           & Slope control of cloud OLR forcing                      & CRETempDiv & Dependence modeled as tanh \\
\enddata
\end{deluxetable*}

\begin{acknowledgements}
FILLET belongs to the CUISINES meta-framework, a Nexus for Exoplanet System Science (NExSS) science working group.
Financial support to R.D. was provided by the Natural Sciences and Engineering Research Council of Canada (NSERC; Discovery Grant RGPIN-2018-05929), the Canadian Space Agency (Grant 18FAVICB21), and the European Research Council (ERC; Consolidator Grant 771620).
P.S. acknowledges the Italian Institute of Oceanography and Applied Geophysics (OGS) and the Italian Inter-University Consortium for Supercomputing (CINECA) for funding his work under the HPC-TRES program (award number 2022-02). J.H.M. acknowledges funding from the NASA Habitable Worlds program under award 80NSSC20K0230.
T.J. Fauchez acknowledges support from the GSFC Sellers Exoplanet Environments Collaboration (SEEC), which is funded in part by the NASA Planetary Science Divisions Internal Scientist Funding Model. R.B. acknowledges support from the NASA Virtual Planetary Laboratory Team through grant number 80NSSC18K0829. Data and scripts for figures 1, 2, 4, and 5 are available at DOI: \dataset[10.5281/zenodo.7563242]{\doi{10.5281/zenodo.7563242}}. 
\end{acknowledgements}

\bibliographystyle{aasjournal}
\bibliography{main}
\end{document}